\documentclass[11pt,a4paper]{article}
\usepackage[margin=2.4cm]{geometry}
\usepackage[T1]{fontenc}
\usepackage{lmodern,microtype,booktabs,amsmath,graphicx,longtable,array}
\newcolumntype{L}[1]{>{\raggedright\arraybackslash}p{#1}}
\usepackage[numbers,sort&compress]{natbib}
\usepackage[hidelinks]{hyperref}
\newcommand{\StudyTasks}{72}

\newcommand{\OrnithNRate}{51.4}

\newcommand{\OrnithRRate}{59.7}

\newcommand{\OrnithDRate}{52.8}

\newcommand{\OrnithFalseAccepts}{25}
\newcommand{\OrnithRequests}{108}

\newcommand{\GemmaNRate}{55.6}

\newcommand{\GemmaRRate}{62.5}

\newcommand{\GemmaDRate}{63.9}

\newcommand{\GemmaFalseAccepts}{11}
\newcommand{\GemmaRequests}{118}

\newcommand{\TotalRecordedRequests}{226}
\newcommand{\NoPublicExamples}{16}
\newcommand{\LowestAdjustedP}{0.125}

\newcommand{\OrnithInitialBlocked}{17}
\newcommand{\OrnithInitialRejected}{18}
\newcommand{\OrnithGateCeilingPass}{55}
\newcommand{\OrnithGateCeilingRate}{76.4}
\newcommand{\OrnithInitialAccepted}{54}

\newcommand{\OrnithNoExampleFA}{9}
\newcommand{\OrnithNoExampleAccepted}{16}
\newcommand{\OrnithWithExampleFA}{4}
\newcommand{\OrnithWithExampleAccepted}{19}
\newcommand{\GemmaInitialBlocked}{9}
\newcommand{\GemmaInitialRejected}{23}
\newcommand{\GemmaGateCeilingPass}{63}
\newcommand{\GemmaGateCeilingRate}{87.5}
\newcommand{\GemmaInitialAccepted}{49}

\newcommand{\GemmaNoExampleFA}{1}
\newcommand{\GemmaNoExampleAccepted}{14}
\newcommand{\GemmaWithExampleFA}{2}
\newcommand{\GemmaWithExampleAccepted}{19}

\newcommand{\OrnithRNSecondsPerNetPass}{7.77}
\newcommand{\OrnithDNSecondsPerNetPass}{56.18}
\newcommand{\GemmaRNExactP}{0.06250}

\newcommand{\GemmaRNSecondsPerNetPass}{35.89}
\newcommand{\GemmaDNSecondsPerNetPass}{27.70}
\newcommand{\GemmaMixedBitOverrides}{126}

\hypersetup{pdftitle={Verification-Gated Redraw and Repair for Local Small Language Models: A Paired Evaluation},pdfauthor={Fedir RYKHTIK}}
\title{Verification-Gated Redraw and Repair for Local Small Language Models: A Paired Evaluation}
\author{Fedir RYKHTIK\\\small ISEN (Institut sup\'erieur de l'\'electronique et du num\'erique)\\\small \texttt{fedir.rykhtik@yncrea.fr}\\\small \url{https://www.fedir.fr}}
\date{October 2026}
\begin{document}
\maketitle
\begin{abstract}
Executable checks can reject incorrect model-generated programs, but rejection
alone does not determine whether fresh sampling or diagnostic repair is preferable.
We present a controlled, resource-bounded conceptual replication and extension of
established recovery policies: a single answer (N), one verification-triggered
redraw (R), and one diagnostic repair (D), for Ornith 1.5 9B and Gemma 4 E4B
served locally in nominal 4-bit MLX format. The frozen evaluation contains \StudyTasks{} tasks: 40 HumanEval+
problems, 16 SQL tasks and 16 shell tasks. Policies share initial responses and
use only public checks for online decisions; every candidate is model-generated.
Each task has one stochastic trajectory per model, so the study does not estimate
variation across independent sampling blocks.
Redraw increases observed final-test success from \OrnithNRate\% to
\OrnithRRate\% for Ornith and from \GemmaNRate\% to \GemmaRRate\% for Gemma.
Diagnostic repair reaches \OrnithDRate\% and \GemmaDRate\%, respectively.
None of four internally prespecified paired contrasts is significant after Holm correction
(minimum adjusted $p=\LowestAdjustedP$). Public checks falsely accept
\OrnithFalseAccepts{} Ornith and \GemmaFalseAccepts{} Gemma candidates, demonstrating
that acceptance is not a correctness guarantee. A post-review audit of initial
answers gives perfect-retry upper bounds of \OrnithGateCeilingRate\% and
\GemmaGateCeilingRate\%, respectively: accepted initial errors cannot trigger
recovery under the fixed gate. This one-block, resource-bounded
study provides a finite-cohort experimental record, not policy equivalence, a general
model ranking or universal reliability.
\end{abstract}

\section{Introduction}
Local small language models (SLMs) can generate executable solutions under
explicit resource limits. A plausible program is not necessarily correct, and
a public example set can accept a program that fails a held-out case. When a
public verifier does reject a response, a recovery policy must decide whether
to sample a fresh solution or ask the model to repair its rejected candidate.

We study two questions: does one verifier-triggered fresh sample improve
selected-answer quality over a single response, and does diagnostic repair
improve on that redraw? We distinguish the answer actually selected by a public
policy from candidate availability, which uses a hidden oracle after collection.
Only the former is deployable under the stated information boundary.

The contribution is a controlled, resource-bounded conceptual replication and
extension of existing recovery comparisons
\citep{liventsev2023autonomous,grishina2025autonomous,arimbur2026tries,kim2026refinement}.
Here, conceptual replication means revisiting the same repair-versus-resampling
question with newly collected responses and a different operational design,
rather than reproducing the cited studies' exact protocols, models or numerical
results. This does not assert that all benchmark tasks are disjoint from those
earlier studies. Specifically, we provide
a paired, internally prespecified comparison of fresh redraw and diagnostic repair
with common requested decoding settings and output caps; extend the task scope
to Python, SQL and shell on two local nominal 4-bit checkpoints; and report selected-answer
outcomes, verifier errors and recorded costs together. Shared initial candidates
and diagnostic/final test separation are established controls, not new algorithms.
Our one-block design samples a single trajectory per task and cannot characterize
run-to-run stability. Historical experiments and development tasks contribute no
observations to the primary endpoint. The study is deliberately narrow and does
not aim to rank SLMs generally.
Figure~\ref{fig:effects} summarizes the observed policy differences; the protocol
and task inventory appear in Appendices~\ref{app:protocol}--\ref{app:tasks}.

\section{Background and related work}
HumanEval evaluates generated programs through execution \citep{chen2021codex};
EvalPlus strengthens test coverage and exposes apparent successes that fail
additional cases \citep{liu2023evalplus}. Feedback-driven refinement and
execution-based debugging are studied in Self-Refine and self-debugging systems
\citep{madaan2023selfrefine,chen2024selfdebug}. Such gains must be distinguished
from obtaining more samples and from the limitations of intrinsic self-correction
\citep{huang2024selfcorrect,olausson2024selfrepair}.

\paragraph{Recent repair--resampling evidence.}
\citet{arimbur2026tries} evaluates iterative execution-feedback repair across seven
hosted models on HumanEval and sanitized MBPP, allowing up to five attempts.
Its direct repair--resampling comparison covers four models on HumanEval and
compares greedy repair with five independent samples at temperature 0.8, reporting
cumulative stopping-test success and resampling pass@5. The paper acknowledges
the decoding difference and imperfect token-budget matching. Its main protocol
does not specify a separate held-out final-grading stage. These outcomes answer
a different selection question from the publicly selected answer graded here:
candidate availability under multiple samples is not itself an online selection
policy. We use the same requested temperature and output cap for R and D, while
retaining the limitation that their actual token costs differ.

\citet{liventsev2023autonomous} study the repair--replace trade-off in SEIDR,
an execution-guided synthesis and debugging framework. On PSB2, intermediate
tree arities of 10 and 100 outperform repair-only and replace-only strategies
in the reported setting. The extension by \citet{grishina2025autonomous} adds
GPT-3.5 and Llama 3-8B, HumanEval-X, repeated runs and lexicase versus tournament
selection. Its reported Llama 3-8B average pass@100 of 84.2\% on HumanEval-C++
is a candidate-availability metric, not our selected-answer success rate.
SEIDR uses multi-round search and candidate ranking under larger budgets;
our comparison permits one retry without population ranking. SEIDR separates
validation and final testing on PSB2, whereas its HumanEval-X debugging loop
exposes all benchmark tests. Thus, repair--replace allocation and separate
final evaluation have precedents, with different search and evaluation contracts.

\paragraph{Controlled one-round refinement.}
\citet{kim2026refinement} compares Direct, feedback-free Self-Refine and
diagnostic-test Self-Debug across seven models and three Python benchmarks:
HumanEval+, MBPP+ and BigCodeBench-Instruct. The protocols share the exact
initial candidate, diagnostic tests are fixed before generation, and final
evaluation uses the complete test suite. Diagnostic tests are a subset of that
suite, not independent final evidence. Self-Refine uses a model review followed
by revision; Self-Debug makes one revision call for every candidate, allowing a
no-change response after a diagnostic pass. Our gate instead stops generation
mechanically on public acceptance and compares repair with a fresh redraw, rather
than with model-generated critique. Kim also reports two additional complete runs,
diagnostic-pass failures left unaddressed, repair/regression patterns, token costs
and overhead per net additional pass. Thus, shared candidates, limited-diagnostic
failures and cost-aware refinement already have direct precedents. Our observed
public false accepts complement that evidence in a smaller, quantized,
cross-language setting; they do not establish a newly discovered failure mechanism.

Table~\ref{tab:recent-work} summarizes the methodological differences. Our smaller
cohort, one sampling block and incomplete collection/runtime reproduction limit the
extension. Different models, checks, decoding settings and endpoints prevent a
cross-paper accuracy or efficiency ranking. D--R compares complete policies:
D changes both diagnostic information and rejected-answer history, so it does
not isolate the causal effect of diagnostic text. Model agreement and
model-generated expected outputs are not treated as independent verification.
\begingroup
\setlength{\tabcolsep}{4pt}
\begin{table}[!t]
\centering\small
\caption{Related recovery and search designs. The 2025 SEIDR article extends the
2023 study. Prior-study facts are author-reported, not independently reproduced
here; budgets and endpoints do not support a cross-paper score or efficiency ranking.}
\label{tab:recent-work}
\begin{tabular}{@{}L{0.15\linewidth}L{0.22\linewidth}L{0.28\linewidth}L{0.29\linewidth}@{}}
\toprule
Study & Setting & Recovery/search design & Checks and endpoint \\
\midrule
Arimbur \citep{arimbur2026tries} & Seven hosted models; Python HumanEval and sanitized MBPP. & Up to five attempts. Four-model repair/resampling comparison: temperature 0 versus 0.8; shared initial responses not specified. & Execution checks stop recovery. Cumulative repair success versus resampling pass@5; separate final suite not specified in the main protocol. \\
\addlinespace
SEIDR, GECCO 2023 \citep{liventsev2023autonomous} & Codex-edit; optional GPT-3 bug summaries. PSB2, Python and C++. & Ranked multi-round search. Arity 1, 10, 100 or effectively infinite; up to 1,000 candidates in the arity comparison (five in prompt experiments). & Validation guides search; separate 2,000-case PSB2 test evaluation. Solved problems and final test pass rate; generated-program cost proxy. \\
\addlinespace
SEIDR, TELO 2025 \citep{grishina2025autonomous} & Extends the initial study with GPT-3.5, Llama 3-8B and HumanEval-X. Python and C++. & Multi-round repair--replace search; tournament/lexicase candidate selection and independent restarts. Larger candidate budgets than one retry. & PSB2: separate tests. HumanEval-X: all tests visible in debugging. Reported pass@100 and across-run candidate availability differ from selected-answer success here. \\
\addlinespace
Kim \citep{kim2026refinement} & Seven hosted/local models; HumanEval+, MBPP+ and BigCodeBench-Instruct. & Shared initial candidate; Self-Refine review/revision versus one Self-Debug revision, including after a diagnostic pass. Three complete runs; model-specific caps. & Diagnostic subset of final tests; complete final-suite grading. Token overhead and cost per net additional pass reported. Diagnostic acceptance does not mechanically skip a call. \\
\addlinespace
This study & Two local nominal 4-bit MLX checkpoints; 40 HumanEval+, 16 SQL and 16 shell tasks. & Paired N/R/D with a shared first response; at most one retry, no population ranking. Requested temperature 0.6; 1,024 output tokens per candidate. One block. & Public checks control stopping; final tests grade selected answers after collection. Actual tokens/time recorded; public offline grading/analysis code (Section~\ref{sec:transparency}). \\
\bottomrule
\end{tabular}
\end{table}
\endgroup

\section{Methods}
\subsection{Models and serving contracts}
The two checkpoints are Ornith 1.5 9B and Gemma 4 E4B in local nominal 4-bit MLX format,
served through LM Studio. One inference workload runs at a time. Both use an
8,192-token context and requested temperature 0.6, top-p 0.95 and top-k 20.
Sampling parameters are archived as requested rather than universally attested.
Ornith uses the OpenAI-compatible chat endpoint with a requested thinking-off
template setting. Gemma uses the native chat endpoint with reasoning explicitly
off. Its native transport has no requested seed field; Ornith requests a seed
label and an offset for recovery. Deterministic seeds and equal actual compute
are not assumed. Saved response statistics report zero reasoning tokens.
Hash-matched checkpoint configs declare default affine 4-bit quantization with
group size 64; Gemma additionally declares \GemmaMixedBitOverrides{} module-level
8-bit overrides. Nominal 4-bit therefore does not mean every tensor is 4-bit.
These static bytes match the pre-collection digests; they do not attest every
effective runtime operation. Appendix~\ref{app:configuration} documents the
recorded configuration, evidence provenance and unknown settings.

The different transport contracts are part of the tested systems. The original
native Ornith pilot request was rejected because reasoning configuration was
unsupported; its error and amendment remain preserved. A versioned development
pilot on six disjoint tasks then completed for each model. These pilot outcomes
were used for preparation only and are excluded from the primary analysis.

\subsection{Frozen task set and information boundary}
The primary set comprises 40 HumanEval+ problems, 16 author-written SQL tasks and
16 author-written shell tasks. There is one sampling block per question and model:
\StudyTasks{} questions per model, not repeated attempts counted as independent
questions. The HumanEval+ subset is a reproducible sample with selection seed
20261025. EvalPlus is pinned to version 0.3.1, HumanEvalPlus to v0.1.10, and the
source digest and task/prompt identities are recorded in the manifest.

Python public checks execute examples explicitly marked as doctests in the
published function docstring, after syntax, import and callable-entry checks.
Literal expected values are compared by value so that quote style and list layout
do not create false failures. A comparison expression without a separate expected
line must evaluate to true. \NoPublicExamples{} of 40 prompts have no runnable
doctest examples and receive syntax/import/callable checks only; prose examples
are not silently converted into synthesized test algorithms. Final grading uses
the official EvalPlus base and additional tests. Public/final tests share benchmark
provenance; independence and absence of training contamination are not claimed.

Each SQL task specifies its schema, output and ordering rules. Queries execute
against public or held-out in-memory SQLite fixtures with read-only authorization.
Shell tasks execute in isolated fixture directories with network/user-directory
access denied, exact stdout/newline comparison, required exit status zero, C locale
and a five-second execution bound. Each practical task has a public fixture and
two final fixtures, including a reordered SQL or additional shell edge case.
Fixture counts are not additional independent sampling units.

The collector reads a serialized public manifest only. Hidden cases, reference
implementations, mutation candidates and final grades are absent from its imports
and diagnostics. Extraction removes code fences and distinguishes a complete
Python function from a body completion; it never supplies missing algorithms or
repairs candidate code. Reference solutions are used only for offline assurance.

\subsection{Assurance and provenance}
Before collection, all 40 official Python canonical completions passed public
and final controls; malformed, wrong-entry-point and semantically wrong controls
failed final grading. The practical suite passed reference/mutation checks and
independent Python fixture-oracle comparisons, SQL row-order transformations and
additional shell edge fixtures. Tests cover whitespace, missing newlines, nonzero
exits, timeouts and blocked reads of user files. This is automated assurance by
the study's implementation process, not independent human review or a formally
verified proof kernel.

The frozen root preserves separate public/private manifests, the analysis plan,
source snapshots, checkpoint hashes and timestamped model protocols. Artifact
keys are case-safe hashes; every attempt stores its raw request/response, task,
prompt, model, block, policy and frozen-protocol association. Dispatch intent is
persisted before sending, and an uncertain interrupted request halts resumption
rather than silently dispatching again. Both models finished collection before
any candidate received a final grade. An integrity audit checks all task/model
cells, shared first-response identities, recorded dispatch states and grade hashes.

\subsection{Policies and budgets}
\begin{description}
\item[N: first answer.] Select the first model response.
\item[R: redraw.] If public verification rejects the first response, request one
fresh sample from the original prompt, without rejected-answer history.
\item[D: diagnostic repair.] After public rejection, ask the model for one repair
using the rejected answer and public diagnostic.
\end{description}
The first raw response is shared across policies. A publicly accepted response
stops recovery; if neither candidate is accepted, the last completed candidate
is retained. Each candidate has a 1,024-output-token cap: N allows one candidate,
R/D at most two. At most three unique requests are needed per task/model because
the first is shared. Request timeout is 120 seconds, and each model has a
1,800-second dispatch window. No optional stopping for significance, hidden-oracle
selection, deterministic answer fallback, adaptive routing or cross-model
escalation is used. Failures remain in the denominator.
The one-retry limit isolates a small additional-call budget; it does not identify
an optimal multi-round or hybrid search strategy.

\subsection{Endpoints and statistical analysis}
The primary endpoint is selected-answer final-test success with all questions
equally weighted. The internally prespecified paired contrasts are R--N and D--R within each
model. Ten percentage points is the stated meaningful effect, not an effect the
study is guaranteed to detect. This fixed, resource-limited cohort has modest
precision; the six-task pilot does not yield a stable power estimate.

Percentile 95\% intervals use 10,000 paired question-bootstrap draws stratified
by the fixed family sizes, with analysis seed 20261025 \citep{efron1993bootstrap}.
These intervals are descriptive, not multiplicity-adjusted. Formal tests use
two-sided exact McNemar p-values with Holm adjustment over the four contrasts
\citep{mcnemar1947,holm1979}. A template-cluster bootstrap is a sensitivity analysis:
SQL and shell use coarse algorithmic groups, while established Python problems
are separate clusters. These groupings cannot remove all task dependence.

For sparse discordances, percentile-bootstrap intervals and exact-test inversion
need not agree. For example, the original Gemma R--N interval excludes zero even
though its raw exact McNemar p-value is \GemmaRNExactP. A positive bootstrap lower
limit is not independent proof of improvement. The main table reports raw and
Holm-adjusted exact p-values side by side.

As post-review sensitivity, we add family-stratified exact-binomial component
bounds with Bonferroni combination across four paired risk differences. The
simultaneous coverage statement is conditional on independent, identically
distributed task pairs within each family; it is not cluster-robust or an estimate
of stochastic sampling-block variation. The construction is intentionally
conservative and does not invert the conditional McNemar test. Its formula,
assumptions and numerical controls are described in Appendix~\ref{app:p1}.
All four bounds include zero and both +10 and -10 percentage points, leaving
the declared meaningful-effect threshold unresolved. The frozen test decisions
and original descriptive bootstrap intervals remain unchanged.

Actual input/output/reasoning tokens, public verification and full-loop times are
recorded. Full-loop time includes request, verification and collector I/O, but
excludes model loading and final grading. There is no deliberate benchmark warmup,
and timing is descriptive of these runs rather than a controlled throughput or
cross-system speed claim. Missing native response IDs and finish reasons remain
null. Cap events are identified by explicit length endings or output counts at
the limit. No exclusions or budget changes followed outcome inspection.
The verifier partitions, conditional recovery and perfect-retry bounds below are
post-review descriptive analyses of saved responses and grades, not prespecified
contrasts. They neither change the original test family nor supply final-test
information to the online policies.

\section{Results}
\begin{figure}[htbp]
\centering\includegraphics[width=\linewidth]{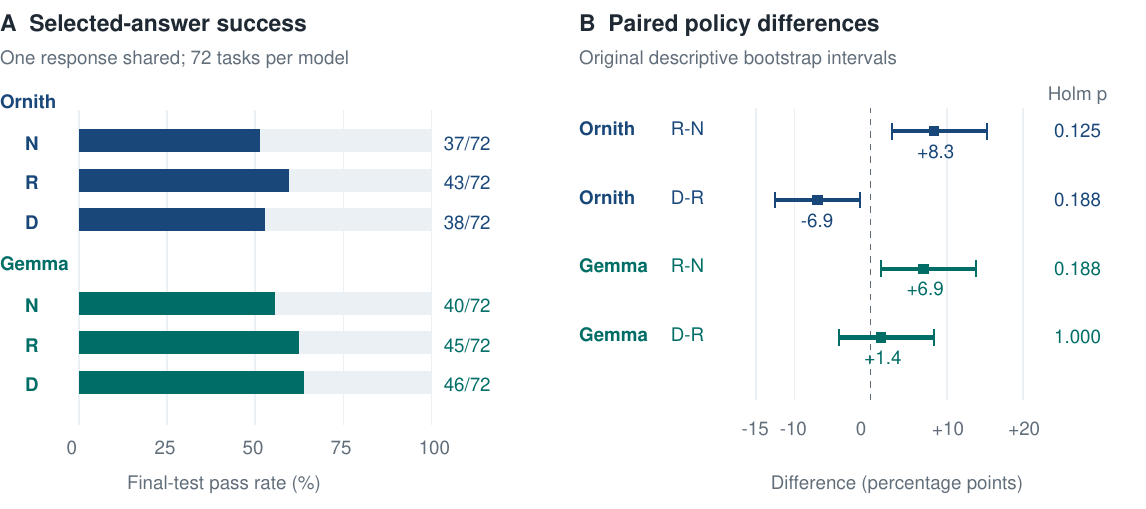}
\caption{Two views of the same frozen outcomes. (A) Selected-answer success under
the three distinct policies, not a time series. (B) Colored bars are original
descriptive, unadjusted 95\% bootstrap intervals. Post-review simultaneous bounds
are reported separately in Appendix~\ref{app:p1}. The right column gives the
internally prespecified Holm-adjusted exact
p-values. Neither a positive estimate nor an unadjusted interval overrides that rule.}
\label{fig:effects}
\end{figure}
\subsection{Selected answers and paired contrasts}
All task/model cells completed. Table~\ref{tab:success} reports selected-answer
success; Table~\ref{tab:effects} gives the internally prespecified contrasts. Fresh redraw
improved the observed counts for both models. Repair was worse than redraw for
Ornith and slightly better for Gemma. No contrast passes the four-test Holm
threshold, so the study establishes neither a policy-superiority claim nor policy
equivalence. The unadjusted bootstrap intervals and adjusted exact tests serve
different purposes and should not be interpreted as the same significance decision.

\begin{table}[ht]
\centering\small\begin{tabular}{lrrr}
\toprule
Model & N & R & D \\
\midrule
Ornith & 37/72 (51.4\%) & 43/72 (59.7\%) & 38/72 (52.8\%) \\
Gemma & 40/72 (55.6\%) & 45/72 (62.5\%) & 46/72 (63.9\%) \\
\bottomrule
\end{tabular}

\caption{Selected-answer final-test success. N, R and D share the same initial
candidate within each model/task. All tasks remain in the denominator.}
\label{tab:success}
\end{table}
\begin{table}[ht]
\centering\small\begin{tabular}{llrrrr}
\toprule
Model & Contrast & Difference (pp) & Bootstrap 95\% CI & Raw $p$ & Holm $p$ \\
\midrule
Ornith & R-N & +8.3 & [+2.8, +15.3] & 0.03125 & 0.1250 \\
Ornith & D-R & -6.9 & [-12.5, -1.4] & 0.06250 & 0.1875 \\
Gemma & R-N & +6.9 & [+1.4, +13.9] & 0.06250 & 0.1875 \\
Gemma & D-R & +1.4 & [-4.2, +8.3] & 1.00000 & 1.0000 \\
\bottomrule
\end{tabular}

\caption{Paired differences in percentage points. Bootstrap intervals are
unadjusted and descriptive; raw two-sided exact McNemar and four-contrast
Holm-adjusted p-values are shown separately.}
\label{tab:effects}
\end{table}

The template-cluster sensitivity retains the observed directions
(Table~\ref{tab:sensitivity}); it does not increase the number of independent
tasks or justify broader extrapolation. Family breakdowns are descriptive
(Table~\ref{tab:families}), with particularly low practical-task success under
the declared exact checks. They are not additional confirmatory hypothesis tests.
\begin{table}[ht]
\centering\small\begin{tabular}{llrrr}
\toprule
Model & Family & N & R & D \\
\midrule
Ornith & HumanEval+ & 22/40 & 25/40 & 23/40 \\
Ornith & SQL & 8/16 & 9/16 & 8/16 \\
Ornith & Shell & 7/16 & 9/16 & 7/16 \\
Gemma & HumanEval+ & 30/40 & 33/40 & 33/40 \\
Gemma & SQL & 4/16 & 5/16 & 5/16 \\
Gemma & Shell & 6/16 & 7/16 & 8/16 \\
\bottomrule
\end{tabular}

\caption{Family-specific selected-answer counts; denominators are distinct tasks.}
\label{tab:families}
\end{table}

\subsection{Verifier failures and candidate availability}
There were \TotalRecordedRequests{} recorded model requests: \OrnithRequests{}
Ornith and \GemmaRequests{} Gemma requests. No HTTP error or unresolved dispatch
state remained. Public checks accepted candidates that later failed final tests
(Table~\ref{tab:verifiers}). Such acceptance blocks both recovery policies, even
when a stronger final check could detect an error. No false public rejection was
observed in the recorded candidate set; that is a finite audit, not verifier proof.
There was one Ornith cap event, and that candidate failed final grading.

\begin{figure}[htbp]
\centering\includegraphics[width=\linewidth]{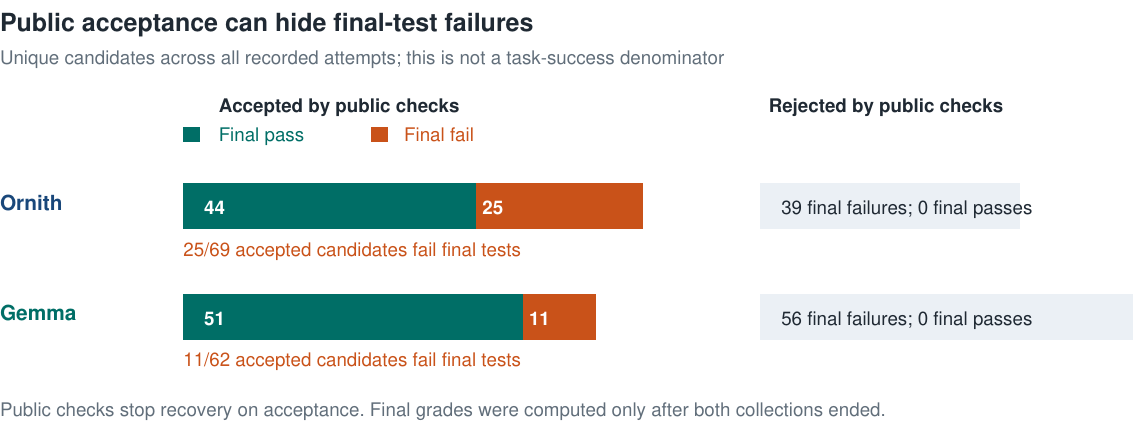}
\caption{The verifier bottleneck. Public acceptance stops recovery, but final
tests reject some accepted candidates (orange). All publicly rejected candidates
also fail final tests in this recorded sample; this does not prove the absence
of false rejections on future candidates. Counts cover unique attempts, including
retries, and therefore differ from the 72-task success denominator.}
\label{fig:verification}
\end{figure}

Candidate availability matched selected-answer success for each policy in this
cohort. This agreement is an observed post-collection property, not permission to
use hidden grades for online selection. Sparse public checks, especially the
Python prompts lacking runnable examples, limit what the deployed policy detects.
The observed acceptance failures show why model agreement or a few passed public
examples cannot establish reliable completion. Figure~\ref{fig:verification}
separates accepted candidates that pass final tests from accepted candidates that
fail them; these are candidate counts, not task-level pass rates.
\begin{table}[ht]
\centering\small\begin{tabular}{lrrrr}
\toprule
Model & False accept & False reject & Requests & Cap events \\
\midrule
Ornith & 25/69 & 0/39 & 108 & 1 \\
Gemma & 11/62 & 0/56 & 118 & 0 \\
\bottomrule
\end{tabular}

\caption{Unique-candidate audit. False-accept denominators are publicly accepted
candidates; false-reject denominators are publicly rejected candidates. Requests
count saved calls, not copies of shared first responses.}
\label{tab:verifiers}
\end{table}

\subsection{Post-review: recovery opportunities and gate-imposed bounds}
The first-answer audit differs from the all-candidate audit above. Among the
initial public accepts, \OrnithInitialBlocked/\OrnithInitialAccepted{} Ornith and
\GemmaInitialBlocked/\GemmaInitialAccepted{} Gemma answers failed final grading.
The gate retained these answers without another model call. Only
\OrnithInitialRejected{} and \GemmaInitialRejected{} initial tasks, respectively,
were rejected and eligible for R/D recovery. The complete initial public/final
decision matrix appears in Appendix~\ref{app:verifier-review}.

\begin{figure}[htbp]
\centering\includegraphics[width=\linewidth]{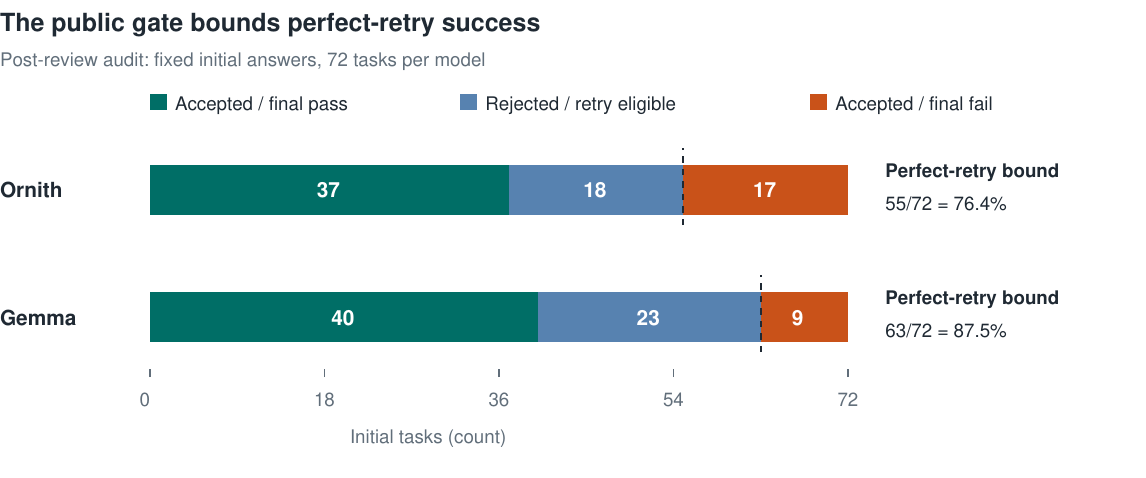}
\caption{Post-review initial-answer decomposition. Green answers are publicly
accepted and pass final tests; blue tasks are rejected and eligible for a retry;
orange answers are accepted but fail final tests. The dashed boundary is the
hypothetical perfect-retry bound for these fixed answers, not an observed policy
score or a claim that every blue task can be solved within the budget.}
\label{fig:ceiling}
\end{figure}

Let $A$ denote initially publicly accepted tasks, $J$ publicly rejected tasks,
and $C$ tasks whose initial answers pass the designated final tests. For fixed
initial responses and the acceptance-stopping rule, a hypothetical perfect retry
can succeed on at most
\[
 U_{\mathrm{gate}}=\frac{|A\cap C|+|J|}{n}
 =1-\frac{|A\cap C^{c}|}{n}.
\]
The formula does not double-count an initially correct but rejected answer.
It gives \OrnithGateCeilingPass/\StudyTasks{}
(\OrnithGateCeilingRate\%) for Ornith and
\GemmaGateCeilingPass/\StudyTasks{} (\GemmaGateCeilingRate\%) for Gemma.
It assumes a permitted-budget successful second candidate on every rejection;
it does not establish that such candidates are realizable, will be generated,
or bound a future task distribution. Figure~\ref{fig:ceiling} shows the partition.

Table~\ref{tab:conditional-recovery} conditions recovery on the rejected initial
tasks. No initially correct answer was rejected in this cohort, so these conditional
successes also equal the repairs of initial final-test failures. Every eligible
task received its permitted R and D attempt; missing outcomes and uncertain
dispatches are zero. Final failures after a completed retry remain failures,
not missing observations. The full retry disposition is retained in
Table~\ref{tab:retry-disposition}. These conditional counts add no new significance
test and do not replace the original full-cohort contrasts.
Because no initially correct answer was publicly rejected, every N success was
accepted and retained unchanged under both R and D. Only initially incorrect
answers entered recovery. Thus N-to-R and N-to-D regressions were structurally
prevented by the observed gate decisions in this cohort. This is not an inherent
monotonicity guarantee: a future false public rejection could replace a correct
initial answer with a worse retry. Nor does it order D against R; neither recovery
policy structurally dominates the other, and Ornith's observed D--R losses remain.
These observations add no D--N test and do not isolate a causal effect of diagnostics.
\begin{table}[ht]
\centering\small\begin{tabular}{lrrrrr}
\toprule
Model & Rejected & R pass & D pass & Missing & Perfect-retry bound \\
\midrule
Ornith & 18 & 6/18 & 1/18 & 0 & 55/72 (76.4\%) \\
Gemma & 23 & 5/23 & 6/23 & 0 & 63/72 (87.5\%) \\
\bottomrule
\end{tabular}

\caption{Post-review recovery conditional on initial rejection. R/D pass entries
use rejected-task denominators; bounds use all 72 tasks. Missing counts are absent
R/D final outcomes; there are also no in-flight dispatches.}
\label{tab:conditional-recovery}
\end{table}

Initial false accepts are not confined to Python: Tables~\ref{tab:initial-strata}
and \ref{tab:candidate-strata} separate task families and initial versus all-attempt
counts. Among accepted initial Python answers, Ornith has
\OrnithNoExampleFA/\OrnithNoExampleAccepted{} false accepts without runnable prompt
examples versus \OrnithWithExampleFA/\OrnithWithExampleAccepted{} with examples.
Gemma has \GemmaNoExampleFA/\GemmaNoExampleAccepted{} versus
\GemmaWithExampleFA/\GemmaWithExampleAccepted{}, respectively. Thus the coverage
association is not consistent across models. The groups differ in task content,
and the gate also selects the retry pool; these are descriptive fractions, not
causal estimates of verifier strength or population error probabilities.

\subsection{Measured costs}
Table~\ref{tab:cost} reports actual mean policy costs, with the shared initial
request charged to each paired policy it serves. Physical collection counts that
request once. D carries additional input history; common output caps therefore
do not imply equal actual compute. The result does not support a general
quality--latency ranking.
\begin{table}[ht]
\centering\small\begin{tabular}{llrrr}
\toprule
Model & Policy & Input tokens & Output tokens & Loop seconds \\
\midrule
Ornith & N & 167.3 & 141.1 & 3.93 \\
Ornith & R & 209.0 & 167.7 & 4.58 \\
Ornith & D & 260.1 & 171.0 & 4.71 \\
Gemma & N & 171.7 & 307.1 & 6.96 \\
Gemma & R & 223.2 & 417.8 & 9.45 \\
Gemma & D & 368.6 & 403.9 & 9.27 \\
\bottomrule
\end{tabular}

\caption{Mean tokens and measured full-loop seconds per task/policy. Loading and
final grading are excluded; no deliberate warmup or deterministic cache control
is claimed.}
\label{tab:cost}
\end{table}

For a more direct cost--gain description, Table~\ref{tab:cost-per-gain} divides the
summed, unrounded policy-minus-N loop seconds and actual input/output tokens by
the net additional final-test passes, not merely the number of attempted or
successful repairs. Redraw costs \OrnithRNSecondsPerNetPass{} additional recorded
loop seconds per net pass for Ornith and \GemmaRNSecondsPerNetPass{} for Gemma;
the corresponding D ratios are \OrnithDNSecondsPerNetPass{} and
\GemmaDNSecondsPerNetPass{}. The shared first answer cancels in these policy
differences but is counted once in physical collection. Ratios are undefined
for zero/negative net gains and unknown token inputs remain unknown.
Small, statistically uncertain net gains make these descriptive ratios unstable;
they do not establish deployment efficiency or a hardware ranking. Exact summed
differences and the formula are in Appendix~\ref{app:p1}.
\begin{table}[ht]
\centering\small\begin{tabular}{llrrrrr}
\toprule
Model & Contrast & \shortstack{Net\\passes} & \shortstack{Added\\seconds} & \shortstack{Seconds/\\net pass} & \shortstack{Input tokens/\\net pass} & \shortstack{Output tokens/\\net pass} \\
\midrule
Ornith & R-N & 6 & 46.64 & 7.77 & 500.0 & 318.7 \\
Ornith & D-N & 1 & 56.18 & 56.18 & 6682.0 & 2148.0 \\
Gemma & R-N & 5 & 179.47 & 35.89 & 741.4 & 1593.2 \\
Gemma & D-N & 6 & 166.18 & 27.70 & 2362.8 & 1160.7 \\
\bottomrule
\end{tabular}

\caption{Post-review incremental recorded costs per net additional final-test
pass versus N. Shared initial requests are charged once within each policy;
loading/final grading are excluded. Ratios have no claimed deployment precision.}
\label{tab:cost-per-gain}
\end{table}
\begin{table}[ht]
\centering\small\begin{tabular}{llr}
\toprule
Model & Contrast & Template-cluster 95\% CI (pp) \\
\midrule
Ornith & R-N & [+3.1, +16.2] \\
Ornith & D-R & [-14.3, -2.6] \\
Gemma & R-N & [+1.4, +13.7] \\
Gemma & D-R & [-4.2, +10.5] \\
\bottomrule
\end{tabular}

\caption{Template-cluster bootstrap sensitivity intervals for paired differences.}
\label{tab:sensitivity}
\end{table}

\section{Discussion and limitations}
Observed redraw gains do not establish a corrected-significance improvement in
this small cohort. Repair does not consistently improve on redraw, and its
contrast also changes rejected-answer context; a diagnosis-specific causal claim
would require an additional control. The appropriate conclusion is uncertainty
under this budget, not that recovery never helps or that the policies are equal.
The additional simultaneous bounds also leave both signs of the 10-point
meaningful effect plausible under their working model. This is imprecision,
not a resolution of practical equivalence or proof that the effect is small.

The post-review bound clarifies a separate architectural limitation: perfect
recovery after rejection cannot fix an incorrect initial answer already accepted
by the unchanged gate. Additional retries on the same rejected branches would
still leave those accepted errors selected. This complements the diagnostic-pass
failures reported by \citet{kim2026refinement}; it is not a new failure mechanism
or an argument that better recovery can never help. The bounds are specific to
these initial responses and checks, rather than intrinsic ceilings of the models.

The custom practical tasks and small fixture sets have author effects, and their
automated expected-result controls may share implementation assumptions. Basic
SQL/text-processing tasks do not represent all software development. Public/final
benchmark overlap, possible training contamination and within-template dependence
limit external validity. One block samples one realized trajectory per task, not
stable repeatability. The two checkpoints are not a representative SLM panel.
Different serving contracts, requested sampler behavior and unknown cache state
complicate cross-model interpretation. Results are scoped to the recorded
checkpoints, task set, checks and budget. No finite observed success rate would
establish completeness on future tasks; this cohort did not reach complete success.

Independent sampling blocks and bounded hybrid repair--redraw policies are useful
next steps. Multi-round repair--replace results motivate that question but do not
establish which allocation works under this local budget. Candidate ranking is a
further design choice; a hybrid policy triggered only by rejection would still
leave publicly accepted initial errors untouched under the same fixed gate.

\section{Transparency and disclosure}
\label{sec:transparency}
This PDF is the standalone report: all primary outcome tables, descriptive figures,
analysis decisions, public-check semantics, task inventory, exact common prompts,
subset IDs and key evidence hashes are included here. Numeric tables, figures and
prose inputs were generated from the frozen evidence, rather than hand-entered.
Appendix~\ref{app:outcomes} includes every paired task outcome and the resampling
rules, enabling independent reconstruction of the statistical comparisons.
It does not include every model response.

An earlier offline replay from frozen source snapshots in a clean temporary directory
reproduced all \TotalRecordedRequests{} candidate grades and the full analysis.
It used the existing pinned evaluator environment and dataset cache, not a fresh
dependency installation or independent review.

Raw requests/responses, policy selections, all saved candidate grades, practical
fixtures, frozen protocols and derived analyses are publicly available at:
\begin{quote}\small
\url{https://github.com/fedir/arxiv-experiments-slm-recover}
\end{quote}
This article references the fixed artifact version
\href{https://github.com/fedir/arxiv-experiments-slm-recover/tree/196a333dd06619eec4ef08a9ed8dd4978879b48e}{commit \texttt{196a333}},
containing \TotalRecordedRequests{} raw candidate records and the corresponding
saved final grades. The repository includes author-data licensing and third-party
notices. It is a versioned GitHub data and offline scientific-code release;
no separate DOI archive has been deposited.

\begin{table}[ht]
\centering\small
\begin{tabular}{@{}L{0.35\linewidth}L{0.59\linewidth}@{}}
\toprule Material & Availability in the cited release \\
\midrule
Requests, responses, selections and saved grades & Public, with recorded provenance and raw-file grade hashes. \\
Practical fixtures, protocols, task outcomes and post-review analyses & Public; held-out practical fixtures were released only after collection. \\
HumanEval final tests & Not bundled; the pinned EvalPlus/HumanEvalPlus component is available upstream. \\
Offline extraction, grader and analysis code & Public under \texttt{scientific/}, with source provenance, documented adapters, dependencies and tests. \\
Original collector and private preflight/pilot logs & Not included; the public package does not reproduce collection or re-audit private configuration evidence. \\
Full runtime and checkpoint weights & Not bundled; identical stochastic replay is not established. \\
\bottomrule
\end{tabular}
\caption{Public data, offline scientific-code and runtime availability at the cited commit.}
\label{tab:availability}
\end{table}

The public package supports provenance checks, formatting-only candidate
extraction, regrading in Python, SQL and shell, and original and post-review statistical
reconstruction. A clean-directory author replay using fresh Python 3.12 dependencies
and a freshly acquired pinned dataset and official-output cache reproduced all
\TotalRecordedRequests{} saved grade records and the full original analysis exactly;
16 release controls passed. Source hashes identify unchanged grading/statistical
functions and documented portability adapters. Outputs are separate from retained
evidence. Validation is on macOS, not an independently performed reproduction or
validation of the Linux execution path. Private configuration telemetry is not
re-audited by the public replay. The complete collection/serving pipeline and
identical stochastic responses remain unreproduced. Hashes identify retained
evidence but do not replace omitted files or establish independent review. The
author retains original source snapshots and additional execution metadata.
\par Public release followed collection;
local protocol timestamps are not public preregistration. Original comparisons
were internally prespecified before collection according to the retained run
records. Protocol and analysis hashes preserve identities, not independently
verified specification timing or priority; the later public release supplies no
such timestamp proof. Review-driven analyses remain retrospective. Code, task assurance,
analysis and writing received AI coding-assistant support. Automated validation
was not independent human review. No checkpoint weights are redistributed.

\paragraph{Funding.}
This study was self-funded by the author and received no external financial support.

\paragraph{Competing interests.}
The author declares no competing interests.

\paragraph{License.}
This article is licensed under
\href{https://creativecommons.org/licenses/by/4.0/}{Creative Commons Attribution 4.0 International (CC BY 4.0)}.

\section{Conclusion}
This conceptual replication compares bounded recovery for local nominal 4-bit models
on \StudyTasks{} tasks in Python, SQL and shell. Redraw improved observed success,
whereas repair-versus-redraw differences were model-dependent. No paired contrast
passed the declared correction. Public false accepts and residual failures
prevented complete success. A post-review audit bounds hypothetical perfect-retry
success at \OrnithGateCeilingRate\% and \GemmaGateCeilingRate\% under the fixed
initial answers and gate; the evidence supports neither policy equivalence
nor universal reliability.
\appendix
\begingroup\small
\section{Protocol and benchmark identity}

\label{app:protocol}

The exact common system instruction was:

\begin{quote}\small

Implement the requested solution. Return only complete code in one code block, including required imports. For Python, define the requested function at top level.

\end{quote}

For R, the original system/user prompt was sampled again without the rejected response. For D, the first response was included as assistant history, followed by the public diagnostic and this exact instruction:

\begin{quote}\small Correct the solution. Return only one code block.\end{quote}

Gemma's native transport serialized the non-system conversation as role-labelled text separated by blank lines; Ornith used structured chat messages. The sole block used requested Ornith seed 142 for the first answer and 1142 for recovery; native Gemma did not receive a seed. Public practical fixtures and their expected outputs were included in the task prompt. Policies were collected in fixed R-then-D order, tasks in Python/SQL/shell order, and Gemma before Ornith; order effects are therefore not randomized.

The 40 selected HumanEval task IDs were:

\begin{quote}\small

5, 8, 10, 11, 13, 15, 17, 18, 19, 21, 25, 34, 36, 50, 55, 57, 58, 59, 65, 69, 82, 83, 86, 91, 93, 97, 99, 104, 107, 108, 112, 113, 118, 120, 123, 126, 132, 146, 149, 161.

\end{quote}

The evaluator used base and additional EvalPlus tests, with its execution guards and default adaptive time limits. Its address-space cap was disabled in this execution environment because the default cap could not be installed; candidate isolation was not a formally verified security boundary.

Table~\ref{tab:identity} gives cryptographic identities for the task/protocol plan and weight files. Public data access is described in Section~\ref{sec:transparency}; not every identified source or weight file is included in that release. Hashes establish identity, not access to omitted material.

\begin{longtable}{@{}L{0.40\linewidth}L{0.55\linewidth}@{}}

\caption{Frozen study identities (SHA-256; each digest wraps after 32 characters).}\label{tab:identity}\\

\toprule Object & Digest \\ \midrule\endfirsthead

\toprule Object & Digest \\ \midrule\endhead

HumanEvalPlus source & \texttt{42526ec0e7d5f3ee0b06d6ced98f8c8b}\newline\texttt{ae3d76519151bfb3d36f79010645bd7f} \\

Public task manifest & \texttt{8303c00c3ed6db12986f4f803cd5d291}\newline\texttt{7f178ee0659b90c088402ba5585dd146} \\

Private practical manifest & \texttt{59ae756ac6df2e9c272322c3bb4ee287}\newline\texttt{0307acc2d7bf22c208ace0698b68e3cf} \\

Analysis plan & \texttt{f1ed7ff09aeac2191f1a9140128aaeb6}\newline\texttt{e9cafa0aa58f708e2925f7b36d62ac9e} \\

Ornith model.safetensors & \texttt{c022a4c4b8de62c75306e2442fd78161}\newline\texttt{ff49bce85fe7abb308c5ad5d04454a26} \\

Gemma model-00001-of-00002.safetensors & \texttt{e58e5ed66324cdc0ce22f51d4d71a9b2}\newline\texttt{db8c24f666278d1b71c3440a18c0f163} \\

Gemma model-00002-of-00002.safetensors & \texttt{5ae9ab6d6e86dd4d8f7eb6957eb7ad15}\newline\texttt{1121431df3855516e89c185824d640cf} \\

\bottomrule\end{longtable}

\clearpage\section{Custom executable task inventory}\label{app:tasks}

The following inventory summarizes the fixed practical specifications. It is not an additional evaluation. Each task had one public fixture and two final fixtures; the latter consisted of an edge fixture plus reordered SQL inserts or a separate shell edge fixture. Public and final expectations were checked against a separate Python fixture oracle. The added SQL order transformation changes physical row order, not the intended result, and does not provide an independent semantic oracle by itself.

\begin{longtable}{@{}L{0.31\linewidth}L{0.65\linewidth}@{}}

\caption{SQL specifications. Each query returns only the requested columns; default binary text collation and read-only execution apply.}\label{tab:sql-inventory}\\

\toprule Schema & Required result \\ \midrule\endfirsthead

\toprule Schema & Required result \\ \midrule\endhead

S1. results(score INTEGER) & Count non-NULL score values. \\

S2. invoices(amount INTEGER) & Return the total invoice amount, treating NULL amounts as zero. \\

S3. products(name TEXT, price INTEGER) & Return every product with the highest price, ordered by name. \\

S4. queue(label TEXT, priority INTEGER) & Return labels ordered by priority ascending, with NULL priorities last, then label. \\

S5. shipments("order" INTEGER) & Return values from the column named order, sorted ascending. \\

S6. sales(id INTEGER) & Return the number of sales rows, including zero for an empty table. \\

S7. customers(id INTEGER, name TEXT); orders(customer\_id INTEGER) & Return customer names with no matching order, sorted by name. \\

S8. routes(origin TEXT, destination TEXT) & Return distinct origin and destination pairs, sorted by origin then destination. \\

S9. games(team TEXT, score INTEGER) & Return each team and its maximum score, sorted by team. \\

S10. jobs(status TEXT) & Return the count of rows whose status is exactly active, ignoring case. \\

S11. events(id INTEGER, day TEXT) & Return event ids dated from 2024-03-01 through 2024-03-31 inclusive, ordered by id. \\

S12. visits(visitor TEXT) & Return visitor names with at least three visits, sorted by name. \\

S13. projects(id INTEGER, name TEXT); issues(project\_id INTEGER, flagged INTEGER) & Return project names having at least one flagged issue, sorted by name. \\

S14. current\_log(id INTEGER); archived\_log(id INTEGER) & Return the total number of rows across current and archived logs. \\

S15. files(filename TEXT) & Return filenames longer than four characters, sorted by filename. \\

S16. readings(temp REAL) & Return the average non-NULL temperature rounded to two decimals. \\

\bottomrule\end{longtable}

\clearpage\begin{longtable}{@{}L{0.31\linewidth}L{0.65\linewidth}@{}}

\caption{Shell specifications. Exact stdout/newlines, zero exit status, C locale and a five-second execution bound apply.}\label{tab:shell-inventory}\\

\toprule Task & Required result \\ \midrule\endfirsthead

\toprule Task & Required result \\ \midrule\endhead

H1. cumulative integer totals & For integers stored one per nonempty line in amounts.txt, print the running total after each line, preserving order. \\

H2. sorted space filenames & Print regular filenames ending in .txt, including names with spaces, sorted one per line. \\

H3. newline record count & Print the number of newline-delimited records in records.txt; an unterminated final record still counts. \\

H4. failed command names & commands.txt contains literal true or false commands. Print only the commands that exit nonzero, in input order. \\

H5. first fields & Print the first colon-separated field of each nonempty line in users.txt. \\

H6. unique casefold words & Print unique words from words.txt case-insensitively, as lowercase, sorted. \\

H7. largest byte file & Print the name of the largest regular file in the current directory; choose the lexicographically greatest full filename on ties. At least one regular file exists. \\

H8. csv active names & From people.csv with header name,status, print active names sorted. \\

H9. line prefix number & Prefix every line of input.txt, including blank lines, with its 1-based number and a colon. \\

H10. deduplicate adjacent lines & Print text.txt after removing only adjacent duplicate lines. \\

H11. matching extension count & Print the count of .log files, including zero when none exist. \\

H12. reverse fields & For each nonempty comma-separated line in pairs.csv, print its two fields in reverse order. \\

H13. nonoverlapping literal count & Print the number of lines in data.txt containing the literal substring a.b (not a regular expression), including zero when none match. \\

H14. last nonempty line & Print the last nonempty line in log.txt after trimming trailing whitespace. \\

H15. tab column sum & Print the integer sum of the second tab-separated column in values.tsv, skipping its header. \\

H16. literal bracket search & Print lines from search.txt containing the literal text [ok]. \\

\bottomrule\end{longtable}

\section{Reading the paired tests}\label{app:discordance}

For each contrast, the exact paired test counts tasks improved versus worsened. These discordances, not repeated calls, determine the p-value. The entries below are generated from the fixed task-level outcomes.

\begin{center}\begin{tabular}{llrr}\toprule Model & Contrast & Improved & Worsened \\ \midrule

Ornith & R-N & 6 & 0 \\

Ornith & D-R & 0 & 5 \\

Gemma & R-N & 5 & 0 \\

Gemma & D-R & 3 & 2 \\

\bottomrule\end{tabular}\end{center}

A positive descriptive point estimate or an unadjusted interval excluding zero does not override the predeclared four-test Holm decision. No adjusted contrast reached the 0.05 threshold; that leaves uncertainty rather than establishing equivalence.

\clearpage\section{Complete paired task outcomes}\label{app:outcomes}

The matrix contains every selected-answer result, including failures: each three-digit code is N/R/D in that order, where 1 means final-test pass and 0 means failure. O is Ornith and G is Gemma. P labels the HumanEval ID; S/H labels the SQL/shell inventory above. For clustering, every P task is its own cluster. SA/SR/SF are SQL aggregation/relational/filter-order groups; HA/HF/HT are shell aggregation/file-selection/text-transform groups. Thus the PDF contains task-level outcomes, rather than only headline aggregates.

\begin{center}\footnotesize\begin{tabular}{llcc@{\hspace{1.5em}}llcc}\toprule Task & Group & O(NRD) & G(NRD) & Task & Group & O(NRD) & G(NRD) \\ \midrule

P5 & P & \texttt{111} & \texttt{111} & P132 & P & \texttt{000} & \texttt{001} \\

P8 & P & \texttt{111} & \texttt{111} & P146 & P & \texttt{111} & \texttt{111} \\

P10 & P & \texttt{000} & \texttt{111} & P149 & P & \texttt{111} & \texttt{111} \\

P11 & P & \texttt{010} & \texttt{111} & P161 & P & \texttt{000} & \texttt{111} \\

P13 & P & \texttt{111} & \texttt{111} & S1 & SA & \texttt{111} & \texttt{000} \\

P15 & P & \texttt{111} & \texttt{111} & S2 & SA & \texttt{111} & \texttt{111} \\

P17 & P & \texttt{000} & \texttt{010} & S3 & SF & \texttt{000} & \texttt{111} \\

P18 & P & \texttt{111} & \texttt{111} & S4 & SF & \texttt{000} & \texttt{000} \\

P19 & P & \texttt{111} & \texttt{111} & S5 & SF & \texttt{111} & \texttt{000} \\

P21 & P & \texttt{111} & \texttt{111} & S6 & SA & \texttt{010} & \texttt{000} \\

P25 & P & \texttt{111} & \texttt{111} & S7 & SR & \texttt{111} & \texttt{011} \\

P34 & P & \texttt{111} & \texttt{111} & S8 & SR & \texttt{000} & \texttt{000} \\

P36 & P & \texttt{111} & \texttt{010} & S9 & SA & \texttt{000} & \texttt{111} \\

P50 & P & \texttt{111} & \texttt{111} & S10 & SF & \texttt{000} & \texttt{000} \\

P55 & P & \texttt{111} & \texttt{111} & S11 & SF & \texttt{111} & \texttt{000} \\

P57 & P & \texttt{111} & \texttt{111} & S12 & SA & \texttt{111} & \texttt{000} \\

P58 & P & \texttt{111} & \texttt{111} & S13 & SR & \texttt{000} & \texttt{111} \\

P59 & P & \texttt{111} & \texttt{111} & S14 & SA & \texttt{111} & \texttt{000} \\

P65 & P & \texttt{000} & \texttt{000} & S15 & SF & \texttt{000} & \texttt{000} \\

P69 & P & \texttt{111} & \texttt{111} & S16 & SA & \texttt{111} & \texttt{000} \\

P82 & P & \texttt{111} & \texttt{111} & H1 & HA & \texttt{111} & \texttt{111} \\

P83 & P & \texttt{000} & \texttt{001} & H2 & HF & \texttt{111} & \texttt{111} \\

P86 & P & \texttt{000} & \texttt{000} & H3 & HA & \texttt{111} & \texttt{000} \\

P91 & P & \texttt{000} & \texttt{000} & H4 & HT & \texttt{111} & \texttt{011} \\

P93 & P & \texttt{000} & \texttt{000} & H5 & HT & \texttt{111} & \texttt{000} \\

P97 & P & \texttt{000} & \texttt{111} & H6 & HT & \texttt{000} & \texttt{111} \\

P99 & P & \texttt{000} & \texttt{000} & H7 & HF & \texttt{010} & \texttt{001} \\

P104 & P & \texttt{111} & \texttt{111} & H8 & HT & \texttt{111} & \texttt{111} \\

P107 & P & \texttt{111} & \texttt{111} & H9 & HT & \texttt{010} & \texttt{000} \\

P108 & P & \texttt{010} & \texttt{011} & H10 & HT & \texttt{000} & \texttt{000} \\

P112 & P & \texttt{111} & \texttt{111} & H11 & HA & \texttt{000} & \texttt{000} \\

P113 & P & \texttt{011} & \texttt{111} & H12 & HT & \texttt{000} & \texttt{111} \\

P118 & P & \texttt{000} & \texttt{111} & H13 & HA & \texttt{000} & \texttt{000} \\

P120 & P & \texttt{000} & \texttt{111} & H14 & HT & \texttt{000} & \texttt{000} \\

P123 & P & \texttt{000} & \texttt{111} & H15 & HA & \texttt{111} & \texttt{111} \\

P126 & P & \texttt{000} & \texttt{111} & H16 & HT & \texttt{000} & \texttt{000} \\

\bottomrule\end{tabular}\end{center}

For each contrast, the analysis resets Python's random.Random to seed 20261025. Read the matrix's left block from top to bottom, then its right block; this is the task order. Family order is Python/SQL/shell and cluster order follows first occurrence. In each of 10,000 draws it samples questions with replacement within each family, preserving family sizes, and then samples template clusters within each family from the same RNG before the next draw. The cluster sensitivity weights family means by 40/72, 16/72 and 16/72. Percentile endpoints are sorted-draw indices floor(0.025(B-1)) and floor(0.975(B-1)), for B=10,000. Exact McNemar tests use twice the lower binomial tail at probability 1/2, capped at one; Holm adjusts across the four specified contrasts.

\clearpage
\begingroup
\setlength{\floatsep}{6pt}
\setlength{\textfloatsep}{6pt}
\setlength{\intextsep}{6pt}
\setlength{\abovecaptionskip}{4pt}
\section{Post-review verifier audit}\label{app:verifier-review}
This descriptive reconstruction reuses original grades checked against raw-file
hashes. It changes neither the original policy decisions nor the prespecified
tests. First-answer results and all-attempt results have different denominators.
\begin{table}[!ht]
\centering\small\begin{tabular}{lrrrr}
\toprule
Model & Accept/pass & Accept/fail & Reject/pass & Reject/fail \\
\midrule
Ornith & 37 & 17 & 0 & 18 \\
Gemma & 40 & 9 & 0 & 23 \\
\bottomrule
\end{tabular}

\caption{Initial public decision crossed with final grade; 72 tasks per model.
Each row partitions every initial answer exactly once.}\label{tab:initial-matrix}
\end{table}
\begin{table}[!ht]
\centering\small\begin{tabular}{llrrr}
\toprule
Model & Stratum & $n$ & FA / accepted & FR / rejected \\
\midrule
Ornith & HumanEval+ & 40 & 13/35 & 0/5 \\
Ornith & SQL & 16 & 1/9 & 0/7 \\
Ornith & Shell & 16 & 3/10 & 0/6 \\
Ornith & Python: no examples & 16 & 9/16 & --- \\
Ornith & Python: examples & 24 & 4/19 & 0/5 \\
Gemma & HumanEval+ & 40 & 3/33 & 0/7 \\
Gemma & SQL & 16 & 2/6 & 0/10 \\
Gemma & Shell & 16 & 4/10 & 0/6 \\
Gemma & Python: no examples & 16 & 1/14 & 0/2 \\
Gemma & Python: examples & 24 & 2/19 & 0/5 \\
\bottomrule
\end{tabular}

\caption{Initial-answer strata: FA is final failure among public accepts; FR is
final pass among public rejections. Python coverage rows partition HumanEval+
and are not additional tasks. A dash means an empty denominator.}
\label{tab:initial-strata}
\end{table}
\begin{table}[!ht]
\centering\small\begin{tabular}{llrrr}
\toprule
Model & Stratum & $n$ & FA / accepted & FR / rejected \\
\midrule
Ornith & HumanEval+ & 50 & 17/43 & 0/7 \\
Ornith & SQL & 30 & 1/10 & 0/20 \\
Ornith & Shell & 28 & 7/16 & 0/12 \\
Ornith & Python: no examples & 16 & 9/16 & --- \\
Ornith & Python: examples & 34 & 8/27 & 0/7 \\
Gemma & HumanEval+ & 54 & 3/39 & 0/15 \\
Gemma & SQL & 36 & 3/9 & 0/27 \\
Gemma & Shell & 28 & 5/14 & 0/14 \\
Gemma & Python: no examples & 20 & 1/16 & 0/4 \\
Gemma & Python: examples & 34 & 2/23 & 0/11 \\
\bottomrule
\end{tabular}

\caption{Recorded candidate attempts, counting each saved call once, including
retries. Identical code from separate calls is not deduplicated. The gate selects
which failures generate extra attempts; these counts are not
independent draws or task-level error probabilities.}\label{tab:candidate-strata}
\end{table}
\begin{table}[!ht]
\centering\small\begin{tabular}{llrrrrr}
\toprule
Model & Policy & Eligible & Final pass & Public reject & Accept/fail & Missing \\
\midrule
Ornith & R & 18 & 6 & 9 & 3 & 0 \\
Ornith & D & 18 & 1 & 12 & 5 & 0 \\
Gemma & R & 23 & 5 & 17 & 1 & 0 \\
Gemma & D & 23 & 6 & 16 & 1 & 0 \\
\bottomrule
\end{tabular}

\caption{Completed retries after initial rejection. In this cohort, final passes,
public rejections and accepted final failures partition the eligible pool.
Missing outcomes and in-flight dispatches are zero. A completed public rejection
is a recorded failure, not an absent outcome.}\label{tab:retry-disposition}
\end{table}
\endgroup

\clearpage
\section{Post-review sparse-pair and cost sensitivity}\label{app:p1}
The original bootstrap intervals and exact tests remain the frozen analysis.
The additional bounds here were chosen after review, with outcomes already known.
For each of M=4 contrasts, H=3 task families and S=2 discordant-cell probabilities,
use a two-sided Clopper--Pearson interval with component error probability
$\alpha_{\rm cell}=0.05/(MHS)=0.05/24$ \citep{clopper1934limits}.
For family h, let b and c be improved and worsened task counts. If their intervals
are $[L_h^+,U_h^+]$ and $[L_h^-,U_h^-]$, the risk-difference bounds are
\[
\left[\sum_h w_h(L_h^+-U_h^-),\;\sum_h w_h(U_h^+-L_h^-)\right],
\quad w_h=n_h/72.
\]
Binomial tails are inverted by bounded bisection, including zero/all-success
boundaries. Implementation controls check closed-form boundaries, a known exact
3/10 reference interval, and finite-sample rational binomial/multinomial coverage
enumerations. Zero discordances retain nonzero-width bounds. A union bound
across 24 component intervals gives simultaneous
coverage of at least 95\% for all four population risk differences under the
working model of independent, identically distributed task pairs within each
family. No independence between cells or contrasts is required. This deliberately
conservative construction is not an inversion of the conditional McNemar test.
It is not cluster-robust: template dependence, nonexchangeable tasks and unmeasured
sampling-block variation limit its interpretation. There is no uncertainty about
the already observed task counts; the working model concerns extrapolation.
\par
\begin{table}[!ht]
\centering\small\begin{tabular}{llrr}
\toprule
Model & Contrast & Improved/worsened & Simultaneous bound (pp) \\
\midrule
Ornith & R-N & 6/0 & [-23.9, +37.6] \\
Ornith & D-R & 0/5 & [-35.5, +24.1] \\
Gemma & R-N & 5/0 & [-24.0, +35.9] \\
Gemma & D-R & 3/2 & [-29.3, +31.5] \\
\bottomrule
\end{tabular}

\caption{Post-review simultaneous bounds under the stated working model. All
include zero and both +10/-10 pp; they resolve neither equivalence nor the
declared meaningful-effect threshold.}\label{tab:simultaneous-bounds}
\end{table}
\par
For policy p, the cost normalization is
\[
G_p=\sum_t(Y_{p,t}-Y_{N,t}),\qquad
E_p=\frac{\sum_t(T_{p,t}-T_{N,t})}{G_p}\quad\text{only if }G_p>0.
\]
Apply the same definition to actual input/output tokens. Unknown tokens propagate
to missing totals, never zero; zero/negative gain has no reported per-gain ratio.
Repair and regression counts are combined as net gain. Shared initial inference
is charged once per policy, but once overall in the physical collection.
\par
\begin{table}[!ht]
\centering\small\begin{tabular}{llrrr}
\toprule
Model & Contrast & Added input tokens & Added output tokens & Added loop seconds \\
\midrule
Ornith & R-N & 3000 & 1912 & 46.638036 \\
Ornith & D-N & 6682 & 2148 & 56.176234 \\
Gemma & R-N & 3707 & 7966 & 179.471287 \\
Gemma & D-N & 14177 & 6964 & 166.182443 \\
\bottomrule
\end{tabular}

\caption{Unrounded task-level cost differences summed before division. Displayed
seconds are rounded only at output.}\label{tab:incremental-totals}
\end{table}
\par
The ratios describe these saved loop costs, not a general latency law. The net
gains are small; the planned contrasts did not establish a corrected-significance
improvement, and no extra D--N hypothesis test was added. The ratios are unstable and
have no claimed deployment precision or confidence interval. Loading and final
grading are excluded. Different serving contracts, fixed ordering, unknown cache
state and uncontrolled background load preclude a cross-model hardware ranking.

\clearpage\section{Recorded execution configuration}\label{app:configuration}

The table distinguishes direct primary records, qualified same-session pilot evidence and unknowns. No current application version or current hardware query was used to backfill the run. Static checkpoint configuration files were retrieved after review only after exact SHA-256 agreement with the pre-collection digests. This verifies their bytes, not every effective runtime operation.

\begin{longtable}{@{}L{0.23\linewidth}L{0.72\linewidth}@{}}

\caption{Configuration and provenance; unknown values remain unknown.}\label{tab:configuration}\\

\toprule Item & Recorded value / limitation \\ \midrule\endfirsthead

\toprule Item & Recorded value / limitation \\ \midrule\endhead

Processor (qualified) & Apple M4 Pro recorded in the same-session pre-confirmation pilot at 2026-10-25T20:24:36+0100; CPU brand/core count were not separately captured at primary preflight. \\

RAM / OS & 24 GiB in both primary preflights; macOS-26.6.2-arm64-arm-64bit. \\

Software & Protocol records Python 3.12.8. LM Studio application version and underlying MLX Python-library versions were not captured. \\

Selected backend & \path{mlx-llm-mac-arm64-apple-metal-advsimd@1.13.1} advertised selected before both runs; not independent binary/build attestation. \\

Checkpoint origins & Ornith: \path{ornith-ai/Ornith-1.5-9B-MLX-4bit}. Gemma: \path{lmstudio-community/gemma-4-E4B-it-MLX-4bit}. Directory/repository identities come from frozen preparation code; upstream commit revisions were not attested. \\

Quantization & Checkpoint configs match pre-collection digests: affine, default 4 bits, group size 64. Ornith declares no module-specific quantization overrides; Gemma declares 126 8-bit module overrides. Nominal 4-bit does not mean every tensor is 4-bit. \\

Context / sampler & Loaded context 8,192, one model instance/parallel slot. Requested temperature 0.6, top-p 0.95, top-k 20, max output 1024; effectiveness of every sampler field is not attested. \\

Reasoning / seeds & Ornith OpenAI-compatible chat: thinking-off requested, seed 142 then 1142. Gemma native chat: reasoning off, no seed field. Saved statistics report zero reasoning tokens; deterministic seeds or activated thinking controls are not established. \\

Cache / warmup & No deliberate warmup; loading excluded from policy times. Prefix-cache state and cross-policy reuse were not controlled/attested; no claim of cache-free or steady-state latency. \\

Power / memory snapshots & AC power in both preflights. Swap used 1716.00 M (sysctl's printed unit) in each. Free-memory counters: Gemma 42\%, Ornith 52\%. These are snapshots, not continuous monitoring. \\

Thermals / background load & Preflight warning logs reported no recorded thermal/performance warnings. Actual temperatures, thermal trajectory and quiet background workload were not independently captured; no controlled hardware speed claim. \\

Limits / order & HTTP timeout 120 s; dispatch window 1800 s/model; public execution 5 s. EvalPlus 0.3.1 defaults, macOS address-space cap disabled. Fixed order: Gemma then Ornith; Python/SQL/shell; R then D. \\

\bottomrule\end{longtable}

\clearpage
Weight-file digests are in Table~\ref{tab:identity}. The following additional digests pin tokenizer, configuration and index files without asserting an upstream repository revision.

\begin{longtable}{@{}L{0.1\linewidth}L{0.33\linewidth}L{0.49\linewidth}@{}}

\caption{Additional pre-collection SHA-256 identities, wrapped after 32 characters.}\label{tab:config-hashes}\\

\toprule Model & File & SHA-256 \\ \midrule\endfirsthead

\toprule Model & File & SHA-256 \\ \midrule\endhead

Ornith & config.json & \texttt{7ddc5bb8c3057edaa0a3337c1721312a}\newline\texttt{8f3d28ad587b77dd6ac075876929d95d} \\

Ornith & model.safetensors.index.json & \texttt{215f9abadc1cea4d465b01c036dffcac}\newline\texttt{5a84c2fb2386777863eff6928b91ed0b} \\

Ornith & tokenizer.json & \texttt{06b9509352d2af50381ab2247e083b80}\newline\texttt{d32d5c0aba91c272ca9ff729b6a0e523} \\

Ornith & tokenizer\_config.json & \texttt{386ba246ba3dfa4d3a21d8bc8712eba3}\newline\texttt{8de558e5ac94c2c90436d8f94d519d5e} \\

Gemma & config.json & \texttt{8b2b956fab049146585c693a08d1cd62}\newline\texttt{5cef8c84d8eb7ea556cf826d0565b92a} \\

Gemma & generation\_config.json & \texttt{d4226bbe3117d2d253ba4609720ba82c}\newline\texttt{6c4ce4627a9a6ae05387c78983ac03de} \\

Gemma & model.safetensors.index.json & \texttt{edfcf22ac78bded1b0518358c2672596}\newline\texttt{f9267d361c433fad7e593f7238bc15cb} \\

Gemma & processor\_config.json & \texttt{1bd0d00776284f369c1eff5fb631e865}\newline\texttt{dfcdca861e0b7d60dbef27fcf37436a8} \\

Gemma & tokenizer.json & \texttt{cc8d3a0ce36466ccc1278bf987df5f71}\newline\texttt{db1719b9ca6b4118264f45cb627bfe0f} \\

Gemma & tokenizer\_config.json & \texttt{16f4a5a617a11af24b4bf47410980098}\newline\texttt{9e7d2b5b1ca22f73a561ec2398308e8d} \\

\bottomrule\end{longtable}

\endgroup
\clearpage
\small
\bibliographystyle{plainnat}
\bibliography{refs}
\end{document}